\documentclass[10pt,letterpaper]{article}
\usepackage{opex3}

\begin{document}

\title{Self consistent, absolute calibration technique for photon number resolving detectors}

\author{A. Avella,$^{1,2}$ G. Brida,$^1$ I. P. Degiovanni,$^1$ M. Genovese,$^1$ M. Gramegna,$^1$
L. Lolli,$^1$ E. Monticone,$^1$ C. Portesi,$^1$ M. Rajteri,$^{1,*}$ M. L. Rastello,$^1$ E.
Taralli,$^1$ P. Traina,$^1$ and M. White$^{1,3}$}

\address{$^1$INRIM, Strada delle Cacce 91, Torino 10135, Italy \\
$^2$ Dipartimento di Fisica Teorica, Universit$\grave{a}$ degli Studi di Torino, Via P. Giuria 1, Torino  10125, Italy\\
$^3$ NPL, National Physical Laboratory, Hampton Road, Teddington, Middlesex TW11 0LW, UK}

\email{$^*$m.rajteri@inrim.it} 



\begin{abstract} Well characterized photon number resolving detectors are a requirement for many applications
ranging from quantum information and quantum metrology to the foundations of quantum mechanics.
This prompts the necessity for reliable calibration techniques at the single photon level.
In this paper we propose an innovative absolute calibration technique for photon number
resolving detectors, using a pulsed heralded photon source based on parametric down conversion. The
technique, being absolute, does not require reference standards and is independent upon the
performances of the heralding detector. The method provides the results of quantum efficiency for
the heralded detector as a function of detected photon numbers. Furthermore, we prove its validity
by performing the calibration of a Transition Edge Sensor based detector, a real photon number
resolving detector that has recently demonstrated its effectiveness in various quantum information
protocols.
\end{abstract}

\ocis{(270.5570) Quantum detectors; (030.5260) Photon counting; (030.5630) Radiometry.} 


\section{Introduction}
Photon number resolving (PNR) detectors are a fundamental tool in
many different fields of optical science and technology \cite{had,had1}
such as quantum metrology (redefinition of the SI candela unit
\cite{qc}), super-resolution \cite{qm}, foundations of quantum
mechanics \cite{mg}, quantum imaging \cite{qi} and quantum
information \cite{qinf,qinf1,qinf2}.

Unfortunately, most conventional single-photon detectors can only distinguish between zero photons
detected (``no-click") and one or more photons detected (``click"). Photon number resolution can be
achieved by spatially \cite{multiSpatial,multiSpatial1} or temporally \cite{multiTemporal,multiTemporal1} multiplexing these
click / no-click detectors. True PNR detection can be achieved only by exploiting detectors
intrinsically able to produce a pulse proportional to the number of absorbed photons. However,
detectors with this intrinsic PNR ability are few \cite{had,had1}, for example photo-multiplier tubes
\cite{burle,burle1,h} and hybrid photo-detectors \cite{NIST}. At the moment, because of their high
detection efficiency, the most promising PNR detectors are the visible light photon counters
\cite{yamamoto,yamamoto1} and transition edge sensors (TESs) \cite{irwin}.

TESs are based on a
superconducting thin film working as a very sensitive thermometer. They are able to discriminate up
to tens of photons per pulse. TESs have recently found important application to quantum information
experiments \cite{TESNIST,TESNIST1,TESNIST2}, demonstrating their huge potential in this field. For practical
application of these detectors it is fundamental they are appropriately characterised. In
particular, one of the most important figures of merit to be characterised is the detection
efficiency, defined as the overall probability of detecting a single photon impinging on the
detector.

For measuring detection efficiency in the photon counting regime, where conventional standards are
cumbersome, an efficient solution is given by Klyshko's absolute calibration technique
\cite{qurad,qurad1,qurad2}, which exploits parametric down conversion (PDC) as a source of heralded single
photons. Despite this technique being suggested in the seventies \cite{Burnham1970,Klyshko1977},
only in recent years has it developed from demonstrational experiments to more accurate
calibrations \cite{Kwiat1994,Dauler1998,Brida2000,Rarity1987}.
 Nowadays, it has been added to the toolbox of primary radiometric techniques for detector calibration, even though it has only been
deeply studied in the case of single-photon click / no-click detectors \cite{Castelletto2000,Ghazi2000,Migdall2002,Polyakov2007,jessica}.

Recently, other techniques for detector calibration, exploiting PDC in the high gain regime, have
been proposed both for the case of analog detectors \cite{ruo1,ruo1_2} and CCDs \cite{ruo2}. Furthermore,
a new technique for the calibration of single photon detectors was proposed exploiting bright PDC
light \cite{worsley09}.  This technique, strongly based on the assumption of a specific detection
model, can in principle be more accurate than the version of Klyshko, but it is important to
underline that the Klyshko's technique, properly developed by the radiometric community, has been
proven accurate at the level of parts in 10$^{3}$ \cite{Polyakov2007,jessica}, i.e. one order of
magnitude more accurate than discussed in Ref. \cite{worsley09}.

We note that the extension of Klyshko's technique to the PNR detection system is quite
straightforward when considering the PNR as a click / no-click single-photon detector.
Nevertheless, this simple application of Klyshko's method is detrimental to the peculiar property
of the PNR detector, i.e. its PNR ability.

By contrast, in this paper we propose and demonstrate an absolute technique for measuring quantum
efficiency, based also on a PDC heralded single photon source, but exploiting all the information
obtained from the output of the PNR detector. As the technique is absolute no reference standards
are required.

We also note this represents the first quantum efficiency measurement of a TES detector exploiting
an absolute technique. Other researchers \cite{Fukuda, Lita} have demonstrated a substitution
method \cite{radiometry} employing a laser beam, calibrated optical attenuators and calibrated
reference standard detectors.

In particular in section \ref{sec:method} we present the theory of our calibration method, while in section \ref{sec:exp}
we present the experimental setup and the results.

\section{Our method}
\label{sec:method}

We used a pulsed PDC based heralded single photon source to illuminate our TES detector. The
typical output of a PNR detector is an histogram representing the relative frequency of detection
events of a certain number of photons. Specifically, we performed two separate measurements, one in
the presence of and one in the absence of heralded photons, obtaining two data histograms. Starting
from these histograms we estimate the probabilities of observing $i$ photons per heralding count in
the presence and in the absence of the heralded photon, $P(i)$ and $\mathcal{P}(i)$ respectively.
Furthermore, we account for the presence of false heralding counts due to stray light and dark
counts.  As $\xi$ is the probability of having a true heralding count (i.e. not due to stray light
and dark counts), the probability of observing no photons on the PNR detector is the sum of the
probability of non-detection of the heralded photons multiplied by the probability of having no
accidental counts in the presence of a true heralding count and the probability of having no
accidental counts in the presence of a heralding count due to stray light or dark counts:
\begin{equation} \label{p0}
P(0)= \xi [(1- \gamma) \mathcal{P}(0)]+ (1-\xi) \mathcal{P}(0),
\end{equation}
hereafter $\gamma$ is the TES ``total" quantum efficiency, i.e. $\gamma = \tau \eta$ where $\tau$
is the optical and coupling losses from the crystal to the fibre end ((a) in Fig. 1) , and $\eta$
is the quantum efficiency of the TES detector. According to Fig. \ref{fig:setup}, we consider the
TES detector as the system from the fibre end (b) to the sensitive area, since this represents the
real detector for applications. This means that $\eta$ accounts also for the losses of the fibre in
the fridge and the geometrical coupling of the light from the fibre to the TES sensitive area.

Analogously, the probability of observing $i$ counts is
\begin{equation} \label{pi}
P(i)= \xi[(1- \gamma) \mathcal{P}(i)+\gamma \mathcal{P}(i-1)]+ (1-\xi)\mathcal{P}(i) ,
\end{equation}
with i=1,2,..., i.e. the sum of the joint probability of non-detection of the heralded photons and
the probability of having $i$ accidental counts, and the joint probability of detection of the
heralded photons and the probability of having $i-1$ accidental counts both in the presence of a
true heralding count, and the probability of having $i$ accidental counts in the presence of a
heralding count due to stray light or dark counts.

From Eq. (\ref{p0}) the efficiency can be estimated as
\begin{equation} \label{eta0}
\gamma_{0}= \frac{\mathcal{P}(0)-P(0)}{\xi  \mathcal{P}(0) },
\end{equation}
while from Eq. (\ref{pi})
\begin{equation} \label{etai}
\gamma_{i}= \frac{P(i)-\mathcal{P}(i)}{\xi  (\mathcal{P}(i-1)-\mathcal{P}(i)) }.
\end{equation}

It is noteworthy to observe that the set of hypotheses in the context of this calibration
technique is similar to the one in Klyshko's technique, i.e. multiphoton PDC events in the time
interval of the order of DET1 temporal resolution (jitter) should be absolutely negligible.
Furthermore, in our case for each value of $i$ we obtain an estimation for $\gamma$ allowing a test
of consistency for the estimation model.

\section{Method implementation and discussion}
\label{sec:exp}
The PNR detector for implementing the proposed calibration technique is a TES based detector,
suitable for broadband response. The TES sensor consists of a superconducting Ti film proximised by
an Au layer \cite{fab}. Such detectors have been thermally and electrically characterised by
impendence measurements \cite{taralli}. The transition temperature of the TES is $T_c$=121 mK with
$\Delta T_c$=2 mK. It is voltage biased \cite{etf} and mounted inside a dilution refrigerator at a
bath temperature of 40 mK. The TES active area is 20 $\mu$m x 20 $\mu$m and is illuminated with a
single mode, 9.5 $\mu$m core, optical fibre. The fibre is aligned on the TES using a
stereomicroscope \cite{lapo}. The distance between the fibre tip and the detector is approximately
150 $\mu$m. The read out is based on a dc-SQUID array \cite{squid}, mounted close to the detector,
coupled to a digital oscilloscope for signal analysis. The obtained energy resolution is $\Delta
E_{FWHM}=$0.4 eV (64 zJ), with a response time of 10.4 $\mu$s \cite{lapo}.

\begin{figure}[tbp]
\par
\begin{center}
\includegraphics[angle=0, width=9cm]{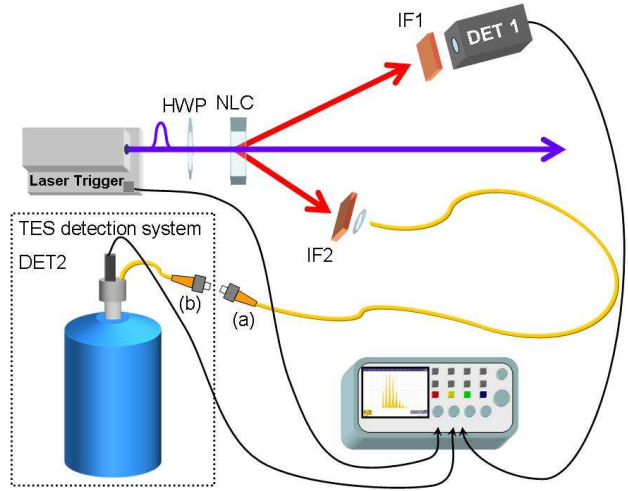}
\caption{Experimental setup: the  heralded single photon sources based on non-collinear degenerate
PDC pumped by 406 nm pulsed laser. The heralding signal from DET1 announces the presence of the
conjugated photon that is coupled in the single mode optical fibre and sent towards the TES based
detector (DET2, identified by the dotted line) starting from the fibre end (b).}\label{fig:setup}
\end{center}
\end{figure}

The calibration is performed using an heralded single photon source based on pulsed non-collinear
degenerate PDC \cite{qurad,qurad1,qurad2} (Fig. \ref{fig:setup}). The heralding photon at 812 nm, emitted at
$3^{o}$ with respect to the pump propagation direction, is spectrally selected by means of an
interference filter 1 nm FWHM (IF1) and detected by a single photon detector DET1 (Perkin-Elmer
SPCM-AQR-14). The heralding signal from DET1 announces the presence of the conjugated photon that
is coupled into the single mode optical fibre and sent towards the TES detector (DET2) after
spectral filtering (IF2 centered at 812 nm with 10 nm FWHM). As usual in quantum efficiency
measurement based on heralding single photon source, the spectral selection is determined by the
trigger detector, while the filter in front of the detector under-test should not reject heralded
photons but it is just inserted to reduce the background counts
[24--37]. 

The pulsed PDC is realised by pumping a type I BBO crystal with a $406$ nm laser, electrically
driven by a train of $80$ ns wide pulses with a repetition rate of 40 kHz. This
low repetition rate is required to avoid pile up effect in the statistics of the measured
counts. In fact it is necessary to use a pulsed heralded single-photon source with a period longer
than the pulse duration of the detector, in order to avoid unwanted photons impinging on the TES
surface before the end of the pulse. If it does not happen, the end effect would be a pile up of
the signal on the tail of the previous detection event with a subsequent extension of the pulse
tail (that can be, somehow, considered the extended dead time of the detector \cite{cdr}).

Despite the pump laser pulse being quite long (80 ns), we note that it is shorter than the temporal
resolution of TESs (time jitter larger than 100 ns). Furthermore, the poor temporal resolution does
not allow the use of small coincidence temporal windows such as the ones used in the typical
coincidence experiments exploiting, for example, Time-to-Amplitude-Converter circuits. One of the
advantages of using a pulsed heralded single photon source is the possibility of evaluating the
events counted by the TES in the presence and absence of an heralding signal, providing an estimate
of the probabilities $P(i)$ and $\mathcal{P}(i)$ in terms of events $C (i)$ and $\mathcal{C}(i)$
counted by the TES. In particular $C (i)$ ($\mathcal{C}(i)$) is the number of events observed by
the TES counting $i$ photons in the presence (absence) of the heralding photon, where $P(i)=
C(i)/\sum_{i} C(i)$ and $\mathcal{P}(i)=\mathcal{C}(i)/\sum_{i} \mathcal{C}(i)$.

\begin{figure}[tbp]
\par
\begin{center}
\includegraphics[angle=0, width=9cm]{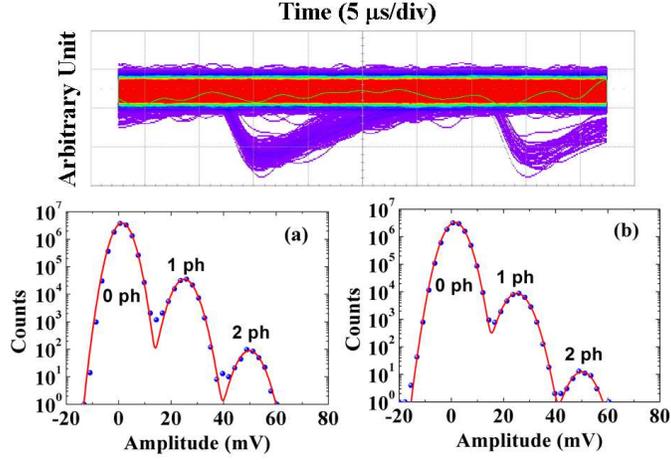}
\caption{Experimental data: oscilloscope screen--shot with traces of the TES detected events. The
group of traces on the left (right) are obtained in the presence (absence) of heralding signals.
Insets (a) and (b) present the histogram of the amplitudes of the pulses in the presence and in the
absence of heralding photons, together with their gaussian fits.} \label{fig:data}
\end{center}
\end{figure}

In Fig. \ref{fig:data} typical traces of the TES detected events observed by the
oscilloscope are shown. The oscilloscope readout is triggered only when both the pump laser trigger
and the heralding detector DET1 clicks are present. The time base is set in order to record on the
trace two subsequent laser pulses. In such a way we are able to measure, on the same trace of the
DET2 pulses, the events containing the heralded photon announced by the contemporary two trigger
signal, i.e. the one corresponding to the laser pulse and the one from DET1 (left pulses in Fig.
\ref{fig:data}), and the subsequent ones not containing the heralded photon (right pulses in Fig.
\ref{fig:data}). By measuring the amplitude of the pulses in this trace we could generate an
histogram where the peaks identify the different number of detected photons corresponding to the
two kind of pulses of Fig. \ref{fig:data}. The insets (a) and (b) are the histograms of the
amplitudes of the pulses in the presence and in the absence of heralding photons, respectively. The
histogram is fitted with gaussian curves $\sum_{i=0}^2[A_i \exp[-(x-x_i)^2/(2\sigma_i^2)]$, where
the fit parameters are the areas $A_i$, the centres $x_i$ and the widths $\sigma_i$ of the Gaussian
curves. The agreement between the experimental data and the fitting is excellent, as stated from
the ratio between the reduced $\chi$-square value and the reduced total sum of square that is lower
than $10^{-4}$. The integrals of the gaussian curves fitted to the histogram peaks provide an
estimate for the parameters $C (i)$ and $\mathcal{C}(i)$. The probability of having true heralding
counts $\xi=0.98793\pm 0.00007$ is obtained as $\xi=1-n_{OFF}/n_{ON}$, where $n_{ON}$ and $n_{OFF}$
are the number of events triggered by the laser pulses and counted by DET1 in the presence and in
the absence of PDC emission, respectively. They correspond in one case to true heralded counts, or
stray light and dark counts, while in the other case only to stray light and dark counts and they
are obtained by means of pump polarization rotation. The PDC extinction provided by the pump
polarization rotation was almost perfect at our pump regime.The measured value for stray light and
dark counts on DET1, are compatible with the values measured with the pump laser blocked before the
crystal. The uncertainty on $\xi$ is evaluated by standard uncertainty propagation on the measured
counts in the presence and in the absence of PDC.

According to Eq.s (\ref{eta0}) and (\ref{etai}), the different estimated values for the ``total"
quantum efficiencies are $\gamma_{0}= (0.709 \pm 0.003) \%$, $\gamma_{1}= (0.709 \pm 0.003) \%$,
and $\gamma_{2}= (0.65 \pm 0.05) \%$. In Table 1 can be found the full analysis of the
uncertainty contributions \cite{GUM}. All the uncertainties are given with coverage factor $k=1$,
obtained from six repeated measurements, each measurement being five hours long, corresponding to
approximately $11 \times 10^6$ heralding counts. The system was very stable during this long run
of measurements. We note that the large uncertainty (derived from standard uncertainty propagation)
in the estimation of $\gamma_{2}$ is essentially due to the poor statistics. For the same reason,
i.e. negligible amount of counted events, it was impossible to obtain estimates of $\gamma$ for
$i>2$. Nevertheless, within its large uncertainty $\gamma_{2}$ is compatible with $\gamma_{0}$ and
$\gamma_{1}$ estimates, which are themselves within very good agreement. This is consistent
with the fact that $\gamma_0$=$\gamma_1$=$\gamma_2$ is expected, since the TES detector has been
recently proved to be a linear detector \cite{tomo}, as generally believed \cite{had,had1}. The
averaged results for the efficiency is $(0.709\pm 0.002)\%$, from a standard weighted mean, where
the uncertainty is calculated accordingly.

\begin{table}
 \caption{Uncertainty contributions in the
measurement of $\gamma_0$, $\gamma1$ and $\gamma2$.  The uncertainty contributions are calculated
according to the well known gaussian propagation of uncertainty formula \cite{GUM}, where the
correlations are accounted for. }
\begin{center}
\begin{tabular}{cccccc}
\hline
Quantity & Value & Standard  &  Unc. & Unc. & Unc.\\
 &  & Uncertainty  &  Contrib.  &  Contrib.  &  Contrib. \\
 &  & & to $\gamma_0 ~(\%)$ & to $\gamma_1 ~(\%)$ & to $\gamma_2 ~(\%)$\\
\hline
$C_0$ & $5.069 ~  10^6$ & $1.4 ~  10^4$ & $-0.003$ & $-0.003$ & $-0.003$\\
   $C_1$ & $5.0200 ~  10^4$ & $200$ & $0.004$ & $0.004$ & $-4 ~ 10^{-5}$\\
   $C_2$ & $118$ & $6$ & $2 ~ 10^{-4}$ & $-2 ~ 10^{-6}$ & $0.05$\\
   $\mathcal{C}_0$ & $5.103 ~ 10^6$ & $1.4  ~  10^4$ & $8 ~ 10^{-4}$ & $8 ~ 10^{-4}$ & $0.003$\\\
   $\mathcal{C}_1$ & $1.4600 ~  10^4$ & $150$ & $-0.003$ & $-0.003$ & $-0.007$\\
   $\mathcal{C}_2$ & $23.9$ & $1.5$ & $-3 ~ 10^{-5}$ & $3 ~ 10^{-7}$ & $-0.02$\\
   $\xi$ & $0.98794$ & $7~ 10^{-5}$ & $-6 ~ 10^{-5}$ & $-6 ~ 10^{-5}$ & $-5 ~ 10^{-5}$\\
   \hline
   $\gamma_0 ~(\%)$ & $0.709$ &  & $0.003$ &&\\
   $\gamma_1 ~(\%)$ & $0.709$ &  && $0.003$ &\\
   $\gamma_2 ~(\%)$ & $0.65$&  &&& $0.05$ \\
\hline
\end{tabular}
\end{center}
\end{table}

In order to validate the proposed technique we compared the efficiencies obtained with those
obtained following the well developed Klyshko's technique \cite{Klyshko1977}. We point out that the
extension of Klyshko's technique to a TES detector is absolutely straightforward. In fact the PNR
ability of the TES detector can be disregarded, it being considered as a click / no-click detector.
Along the guidelines of Ref.s \cite{Polyakov2007, jessica} and from the same experimental data used
for the evaluation of the $\gamma_i$'s we evaluate the ``total'' efficiency in the case of the
Klyshko's technique, obtaining $\gamma_{Klyshko}=(0.707 \pm 0.003) \%$. The result is in perfect
agreement with that obtained from the proposed new technique as implemented in the work reported
here.

The evaluation of the total efficiency $\gamma$, instead of $\eta$, allows us a better comparison
between the results obtained from the two techniques, as the additional independent measurement of
$\tau$ is common to the two techniques. For this reason, in the context of the comparison, it only
provides an additional and somewhat misleading common uncertainty contribution \cite{footnote}.

We notice that the value of the measured efficiency is rather small with respect to results
presented in the literature, e.g. \cite{Fukuda,Lita}. However, the measured values are absolutely
consistent within the context of our experimental setup. Note that the TES sensitive area is not
optimised for detection efficiency at a specific wavelength. On the basis of the material used the
expected efficiency of the TES is approximately $49\%$, while the parameter $\tau$ is estimated to
be $10 \%$. The geometrical and optical losses inside the refrigerator contribute to lower the
value of $\eta$ to $7\%$.

\section{Conclusions}

In conclusion, we have proposed and demonstrated a new absolute calibration technique, applicable
to TES detectors, with an estimated relative uncertainty of $10^{-3}$, that does not rely on
reference standards. Considering the importance of PNR detectors in advancing quantum technologies,
this result represents an important step in their precise characterisation, paving the way to
metrological applications of this absolute method.

\section*{Acknowledgments}

This work was supported by the European Community's Seventh Framework Program, ERA-NET Plus, under
Grant Agreement No. 217257.


\begin{thebibliography}{99}

\bibitem{had} R. H. Hadfield, ``Single-photon detectors for optical quantum information applications,'' Nature Photon. {\bf 3,} 696--705 (2009) \textit{and ref.s therein.}

\bibitem{had1} C. Silberhorn, ``Detecting quantum light,'' Contemp. Phys. \textbf{48,} 143--156 (2007) \textit{and ref.s therein.}

\bibitem{qc} J. C. Zwinkels, E. Ikonen, N. P. Fox, G. Ulm, and M. L. Rastello, ``Photometry, radiometry and 'the candela': evolution in the classical and quantum world,'' Metrologia \textbf{47,} R15-R32 (2010).

\bibitem{qm}Y. Gao, P. M. Anisimov, C. F. Wildfeuer, J. Luine, H. Lee, and J. P. Dowling, ``Super-resolution at the shot-noise limit with coherent states and photon-number-resolving detectors,'' J.Opt. Soc. Am. B \textbf{27}, A170--A174
(2010).

\bibitem{mg} M. Genovese, ``Research on hidden variable theories: A review of recent progresses,''  Phys. Rep. \textbf{413}, 319-396  (2005) \textit{and ref.s therein.}


\bibitem{qi} G. Brida, M. Genovese, and I. Ruo Berchera, ``Experimental realization of sub-shot-noise quantum imaging,''  Nature Photon. \textbf{4}, 227-230 (2010).

\bibitem{qinf} T. Laenger, and  G. Lenhart, ``Standardization of quantum key distribution and the ETSI standardization initiative ISG-QKD,'' New J. Phys. \textbf{11}, 055051 (2009) \textit{and ref.s therein.}

\bibitem{qinf1} J. L. O'Brien, A. Furusawa, and J. Vu\v{c}kovi\'{c}, ``Photonic quantum technologies,'' Nature Photon. \textbf{3}, 687--695 (2009) \textit{and ref.s therein.}

\bibitem{qinf2} N. Gisin and R. Thew, ``Quantum communication,''  Nature
Photon. \textbf{1}, 165--171 (2007) \textit{and ref.s therein.}

\bibitem{multiSpatial} L. A. Jiang, E. A. Dauler, and J. T. Chang,  ``Photon-number-resolving detector with 10 bits of resolution,'' \pra \textbf{75}, 062325
(2007).

\bibitem{multiSpatial1} A. Divochiy, F. Marsili, D. Bitauld, A. Gaggero, R. Leoni, F. Mattioli, A. Korneev, V. Seleznev, N. Kaurova, O. Minaeva, G. Gol'tsman, K. G. Lagoudakis, M. Benkhaoul, F. Lévy, and A. Fiore, ``Superconducting nanowire photon-number-resolving detector at telecommunication wavelengths,'' Nature Photon. \textbf{2}, 302--306 (2008).

\bibitem{multiTemporal} D. Achilles, C. Silberhorn, C. Sliwa, K. Banaszek, and I. A.  Walmsley, ``Fiber-assisted detection with photon number resolution,'' \ol \textbf{28}, 2387-2389 (2003).

\bibitem{multiTemporal1} M. J. Fitch, B. C. Jacobs, T. B. Pittman, and J. D. Franson, ``Photon-number resolution using time-multiplexed single-photon detectors,'' \pra \textbf{68}, 043814 (2003).

\bibitem{burle} G. Zambra, M. Bondani, A. S. Spinelli, F. Paleari, and A. Andreoni, ``Counting photoelectrons in the response of a photomultiplier tube to single picosecond light pulses,'' Rev. Sci. Instrum. \textbf{75}, 2762
(2004).

\bibitem{burle1} M.Bondani, A. Allevi, and A. Andreoni, ``Light Statistics by Non-Calibrated Linear Photodetectors,''  Advanced Science Letters \textbf{2}, 463--468 (2009).

\bibitem{h} G. A. Morton, RCA Rev. \textbf{10}, 525 (1949).

\bibitem{NIST} M. Ramilli, A. Allevi, V. Chmill, M. Bondani, M. Caccia, and A. Andreoni, ``Photon-number statistics with silicon photomultipliers,'' \josa B \textbf{27}, 852-862 (2010).

\bibitem{yamamoto} J. Kim, S. Takeuchi, Y. Yamamoto, and H. H. Hogue, ``Multiphoton detection using visible light photon counter,'' \apl, \textbf{74}, 902 (1999).

\bibitem{yamamoto1} E. Waks, K. Inoue, W. D. Oliver, E. Diamanti, and Y. Yamamoto, ``High-efficiency photon-number detection for quantum information processing,'' IEEE J. Sel. Top. Quantum Electron \textbf{9}, 1502--1511 (2003).

\bibitem{irwin} K. D. Irwin and G. C. Hilton, ``Transition-Edge Sensors,'' in {\it Cryogenic Particle Detection (Topics Appl. Phys. Vol. 99)}, C. Enss eds., (Springer-Verlag, Berlin, 2005), pp. 63--149.

\bibitem{TESNIST}
A. J. Pearlman, A. Ling, E. A. Goldschmidt, C. F. Wildfeuer, J. Fan, and A. Migdall, ``Enhancing image contrast using
coherent states and photon number resolving detectors,'' \opex \textbf{18}, 6033--6039 (2010).

\bibitem{TESNIST1} T. Gerrits, S. Glancy, T. S. Clement, B. Calkins, A. E. Lita, A. J. Miller, A. L. Migdall, S. W. Nam, R. P. Mirin, and E. Knill, ``Generation of optical coherent-state superpositions by number-resolved photon subtraction from the squeezed vacuum,''
\pra \textbf{82}, 031802 (2010).

\bibitem{TESNIST2} K. Tsujino, D. Fukuda, G. Fujii, S. Inoue, M. Fujiwara, M. Takeoka, and M. Sasaki, ``Sub-shot-noise-limit discrimination of on-off
keyed coherent signals via a quantum receiver with a superconducting transition edge sensor,'' \opex \textbf{18}, 8107--8114 (2010).

\bibitem{qurad} A. Migdall, ``Correlated-photon metrology without absolute standards,'' Phys. Today \textbf{52}, 41--46 (1999) \textit{and ref.s therein.}

\bibitem{qurad1} G. Brida, M. Genovese, and M. Gramegna, ``Twin-photon techniques for photo-detector calibration,'' Laser Physics Lett. \textbf{3}, 115--123 (2006) \textit{and ref.s therein.}

\bibitem{qurad2} S. V. Polyakov, and A. L. Migdall, ``Quantum radiometry,'' \jmo \textbf{56}, 1045--1052 (2009) \textit{and ref.s therein.}

\bibitem{Burnham1970} D. C. Burnham and D. L. Weinberg, ``Observation of Simultaneity in Parametric Production of Optical Photon Pairs,'' \prl \textbf{25}, 84--87 (1970).

\bibitem{Klyshko1977} D. N. Klyshko, ``Utilization of vacuum fluctuations as an optical brightness standard,''  \sjqe \textbf{7}, 591 (1977).

\bibitem{Kwiat1994} P. G. Kwiat, A. M. Steinberg, R. Y. Chiao, P. H. Eberhard, and M. D. Petroff, ``Absolute efficiency and time-response measurement of single-photon detectors,'' \ao \textbf{33}, 1844--1853 (1994).

\bibitem{Dauler1998} E. Dauler, A. L. Migdall, N. Boeuf, R. U. Datla, A. Muller, and A. Sergienko, ``Measuring absolute infrared spectral radiance with correlated photons: new arrangements for improved uncertainty and extended IR range,'' Metrologia \textbf{35}, 295 (1998).

\bibitem{Brida2000} G. Brida, S. Castelletto, I. P. Degiovanni, M. Genovese, C. Novero, and M. L. Rastello, ``Towards an uncertainty budget in quantum-efficiency measurements with parametric fluorescence,'' Metrologia \textbf{37}, 629 (2000).

\bibitem{Rarity1987} J. G. Rarity, K. D. Ridley, and P. R. Tapster, ``Absolute measurement of detector quantum efficiency using parametric downconversion,'' \ao \textbf{26}, 4616--4619 (1987).

\bibitem{Castelletto2000} S. Castelletto, I. P. Degiovanni, and M. L. Rastello, ``Evaluation of statistical noise in measurements based on correlated photons,'' \josa B \textbf{19}, 1247--1258 (2002).

\bibitem{Ghazi2000} A. Ghazi-Bellouati, A. Razet, J. Bastie, M. E. Himbert, I. P. Degiovanni, S. Castelletto, and M. L. Rastello, ``Radiometric reference for weak radiations: comparison of methods,''  Metrologia \textbf{42}, 271 (2005).

\bibitem{Migdall2002} A. L. Migdall, S. Castelletto, I. P. Degiovanni, and M. L. Rastello, ``Intercomparison of a Correlated-Photon-Based Method to Measure Detector Quantum Efficiency,'' \ao \textbf{41}, 2914-2922 (2002).

\bibitem{Polyakov2007} S.V. Polyakov and A.L. Migdall, ``High accuracy verification of a correlated-photon-based method for determining photoncounting detection efficiency,'' \opex \textbf{15}, 1390--1407 (2007).

\bibitem{jessica} J. Y. Cheung, C. J. Chunnilall, G. Porrovecchio, M. Smid, E. Theocharous, ``Low optical power reference detector implemented in the validation of two independent techniques for calibrating photon-counting detectors,'' \opex (\textit{submitted}).

\bibitem{ruo1} G. Brida, M. Genovese, I. Ruo-Berchera, M. Chekhova, and A. Penin,  ``Possibility of absolute calibration of analog detectors by using parametric downconversion: a systematic study,'' \josa B \textbf{23}, 2185--2193 (2006).

\bibitem{ruo1_2} G. Brida, M. Chekhova, M. Genovese, and I. Ruo-Berchera, ``Analysis of the possibility of analog detectors calibration by exploiting stimulated parametric down conversion,'' \opex \textbf{16}, 12550--12558 (2008).

\bibitem{ruo2} G. Brida, I. P. Degiovanni, M. Genovese, M. L. Rastello, and I. Ruo Berchera, ``Detection of multimode spatial correlation in PDC and application to the absolute calibration of a CCD camera,'' \opex \textbf{18}, 20572-20584 (2010).

\bibitem{worsley09} A. P. Worsley, H. B. Coldenstrodt-Ronge, J. S. Lundeen, P. J. Mosley, B. J. Smith, G. Puentes, N. Thomas-Peter, and I. A. Walmsley,  ``Absolute efficiency estimation of photon-number-resolving detectors using twin beams,'' \opex {\bf 17},
4397--4411 (2009).

\bibitem{Fukuda} D. Fukuda, G. Fujii, T. Numata, K. Amemiya, A. Yoshizawa, H. Tsuchida, H. Fujino, H. Ishii, T. Itatani, S. Inoue, and T. Zama, ``Titanium-based transition-edge photon number resolving detector with 98\% detection efficiency with index-matched small-gap fiber coupling,'' \opex \textbf{19}, 870-875 (2011).

\bibitem{Lita} A.E. Lita, A. J. Miller,and S. W. Nam, ``Counting near-infrared single-photons with 95\% efficiency,'' \opex \textbf{16}, 3032--3040 (2008).

\bibitem{radiometry} A. C. Parr, R. U. Datla, J. L. Gardner, \textit{Optical Radiometry} (Elsevier Academic Press, Amsterdam 2005)

\bibitem{fab} C. Portesi, E. Taralli, R. Rocci, M. Rajteri, and E. Monticone, ``Fabrication of Au/Ti TESs for Optical Photon Counting,'' J. Low Temp. Phys. {\bf 151}, 261--265 (2008).

\bibitem{taralli} E. Taralli, C. Portesi, L. Lolli, E. Monticone, M. Rajteri, I. Novikov, and J. Beyer, ``Impedance measurements on a fast transition-edge sensor for optical and near-infrared range,''  Supercond. Sci. Technol. {\bf 23}, 105012 (2010).

\bibitem{etf} K. D.  Irwin, ``An application of electrothermal feedback for high resolution cryogenic particle detection,''  \apl {\bf 66},  1998 (1995).

\bibitem{lapo} L. Lolli, E. Taralli, C. Portesi, D. Alberto, M. Rajteri, and E. Monticone, ``Ti/Au Transition-Edge Sensors Coupled to Single Mode Optical Fibers Aligned by Si V-Groove,''  IEEE Trans. Appl. Supercond.  {\bf 21} 215--218 (2011).

\bibitem{squid} D. Drung, C. Assmann, J. Beyer, A. Kirste, M. Peters, F. Ruede, and T. Schurig, ``Highly Sensitive and Easy-to-Use SQUID Sensors,'' IEEE Trans. Appl. Supercond. {\bf 17}, 699--704 (2007).

\bibitem{cdr} S. Castelletto, I. P. Degiovanni, M. L. Rastello, ``Theoretical aspects of photon number measurement,'' Metrologia {\bf 37}, 613--616 (2000).

\bibitem{GUM} Guide to the Expression of Uncertainty in Measurement, ISO (1995).

\bibitem{tomo} G. Brida, L. Ciavarella, I. P. Degiovanni, M. Genovese, L. Lolli, M. G. Mingolla, F. Piacentini, M. Rajteri, E. Taralli, M. G. A. Paris, ``Full quantum characterization of superconducting photon counters,'' http://arxiv.org/pdf/1103.2991.

\bibitem{footnote} Incidentally, if one wants to provide a precise estimate of the naked TES based
detector quantum efficiency $\eta$ it is necessary a careful estimation of the optical
transmittance $\tau$, accounting for the coupling efficiency in the optical fiber and the optical
losses in the non-linear crystal. According to the results of Ref.s [S.V. Polyakov, A.L. Migdall,
\opex \textbf{15}, 1390 (2007); J. Y. Cheung \textit{et al.}, \ao
(\textit{submitted})], one could provide an estimate of this parameter with a less than 1$\%$
uncertainty.

\end{thebibliography}
\end{document}